\documentclass{article}

\usepackage{PRIMEarxiv}

\usepackage[utf8]{inputenc} 
\usepackage[T1]{fontenc}    
\usepackage{hyperref}       
\usepackage{url}            
\usepackage{booktabs}       
\usepackage{amsfonts}       
\usepackage{nicefrac}       
\usepackage{microtype}      
\usepackage{lipsum}
\usepackage{fancyhdr}       
\usepackage{graphicx}       
\graphicspath{{media/}}     
\usepackage{mathtools}

\title{Causal Hierarchy in the Financial Market Network -- Uncovered by the Helmholtz-Hodge-Kodaira Decomposition
}

\author{
  Tobias Wand \\
  University of Münster, Germany \\
  Rissho University, Kumagaya, Japan \\
  \texttt{t\_wand01@uni-muenster.de} \\
   \And
  Oliver Kamps \\
  University of Münster, Germany \\
  \texttt{okamp@uni-muenster.de} \\
   \And
  Hiroshi Iyetomi \\
  Rissho University, Kumagaya, Japan \\
  Canon Institute for Global Studies, Tokyo, Japan\\
  \texttt{hiyetomi@ris.ac.jp} \\
}

\begin{document}
\maketitle

\begin{abstract}
Granger causality can uncover the cause and effect relationships in financial networks. However, such networks can be convoluted and difficult to interpret, but the Helmholtz-Hodge-Kodaira decomposition can split them into a rotational and gradient component which reveals the hierarchy of Granger causality flow. Using Kenneth French's business sector return time series, it is revealed that during the Covid crisis, precious metals and pharmaceutical products are causal drivers of the financial network. Moreover, the estimated Granger causality network shows a high connectivity during crisis which means that the research presented here can be especially useful to better understand crises in the market by revealing the dominant drivers of the crisis dynamics.
\end{abstract}


\section{Introduction} 

One of the most important messages in many introductory lectures to statistics is that correlation does not imply causation \cite{Aldrich1995}. However, it begs the question: What, then, is causality? And how can it be quantified? One of the first and most widespread attempts to formalise causality was proposed by Granger~\cite{GrangerCausality}. Granger causality takes the time ordering into account as the cause needs to happen before the effect. The field of causal inference has developed several tools to probe time series data for causal interactions~\cite{CausalPrimer} and has been used to analyse dynamical systems~\cite{Causality_Attractor_Lorenz,Causality_Ecosystems_Attractor}. Especially for high dimensional multivariate time series, it is difficult to infer the network of causality because one has to carefully distinguish between the different possible causal drivers~\cite{Causal_Inference_Optimal_Entropy,Aste_DiMatteo_2017,GrangerMultivariate_Siggiridou,Siggiridou_Comparison_2019}.\\
While such networks can be convoluted and difficult to interpret, especially if they contain cyclic substructures~\cite{Iskrzyski2021}, the Helmholtz-Hodge-Kodaira Decomposition (HHKD) can disentangle them. As a reformulation of Helmholtz-Hodge field theory for discrete graphs, the HHKD can split a directed network into a cyclic and gradient-based graph \cite{Johnson_HodgeTutorial2013,Strang_PhDThesis}. The latter will then provide a ranking of all nodes according to whether they are upstream or downstream. The application of methods and tools from physics to economic and financial systems is known as econophysics~\cite{MantegnaStanleyBook} and the HHKD has been used in this field to understand the dynamics of cryptocurrencies~\cite{Fujiwara2021} as well as the network of shared ownership of companies~\cite{Kichikawa2021}. Capturing economic and financial interactions in a network has been a standard approach of econophysics and complexity science in the past \cite{WorldTradeWeb_PhysRevE.68.015101,HandbookGraphs,LargeScaleStructure} and causal inference has been applied to such networks of companies or countries~\cite{CausalEntropyFinancialNetwork,GrangerCausality_FinancialNetworks,StanleyCausality}.\\
This article analyses time series data of business sectors from~\cite{FrenchData} to investigate the Granger causality between different sectors of the economy. Using the HHKD on this Granger network then reveals if a sector is rather driven by other sectors or if it is a causal driver of the whole system. First, this article will present the database, the algorithm from~\cite{GrangerMultivariate_Siggiridou} to estimate Granger causality networks and the HHKD for graphs in section~\ref{sec:hhkd_materials_methods}. The results of the HHKD will then be presented for different time periods in section~\ref{sec:hhkd_results} before interpreting them and discussing further extensions to this research in section~\ref{sec:hhkd_discussion}.

\section{Materials and Methods}
\label{sec:hhkd_materials_methods}
\subsection{Data}
\label{sec:DataFrench}
To analyse the interactions between different sectors of the economy, we use the database of Ken French which contains the return time series of representative portfolios for 49 different business sectors \cite{FrenchData}. These portfolios are constructed as value-weighted averages of all stocks in a business sector listed on NYSE, AMEX and NASDAQ and the data consists of the daily returns $R^{(i)}_t = \left(P^{(i)}_t - P^{(i)}_{t-1}\right)/P^{(i)}_{t-1}$ of these portfolios' prices. The calculation of returns ensures the stationarity of the data and this database is updated continuously with further details on the data curation given in \cite{Fama2023}. Although the assignment of companies to a sector was done manually, a comparative study with modern statistical tools shows the high agreement between French's classification and data-driven methods \cite{Chan2007}. While the data was originally used for capital asset pricing modelling in \cite{FAMA1997153}, it has found numerous applications in various fields of economic and financial research as a data resource (see \cite{Chan2007} for an overview).

\subsection{Granger Causality}
The intuition behind Granger causality is that the cause $X$ should happen before the effect $Y$ and that knowing the cause should improve the future prediction of the effect. The latter can be measured by fitting autoregressive linear models with and without $X$ and comparing their accuracy. By including possible alternative causes $Z$ for $Y$, this concept is extended to the conditional Granger causality \textit{CGC}. First, a full model is trained that measures how well the past of $X$, $Y$ and background variables $Z$ predict the future of $Y_{t+1}$ via
    \begin{equation}
    \label{eq:ConditionalGranger_FullModel}
        Y_{t+1} = \sum_{\tau = 0}^{\tau_{\max}} \left( \alpha_\tau Y_{t-\tau} + \beta_\tau X_{t-\tau} + \gamma_\tau Z_{t-\tau}   \right) + \epsilon 
    \end{equation}
with i.i.d. Gaussian errors $\epsilon \sim \mathcal{N}(0,\sigma_F^2)$ and a maximum time lag $\tau_{\max}$ to limit how much of the past should be considered for predicting $Y_{t+1}$. Note that $Z$ might contain more than one background variable $Z = \left( Z^{(1)},\dots,Z^{(s)}\right)$ with $\gamma_\tau \in \mathbb{R}^s$. Then, a reduced model is trained without the proposed cause $X$ as 
\begin{equation}
    \label{eq:ConditionalGranger_RedModel}
        Y_{t+1} = \sum_{\tau = 0}^{\tau_{\max}} \left( \alpha^\prime_\tau Y_{t-\tau} + \gamma^\prime_\tau Z_{t-\tau}   \right) + \epsilon
\end{equation}
with $\epsilon \sim \mathcal{N}(0,\sigma_R^2)$. The conditional Granger causality of $X$ on $Y$ is then given by how much the reduced model's variance increases compared to the full model and is defined as
\begin{equation}
    \label{eq:CGC}
    CGC_{X\rightarrow Y} = \log{\frac{\sigma_R^2}{\sigma_F^2}}
\end{equation}
to measure how much $X$ causes $Y$.\\

For multivariate data with many time series, the estimation of the full model in \eqref{eq:ConditionalGranger_FullModel} can easily fall into the regime of overfitting \cite{OverfittingBabyak}. Hence, it is of paramount importance to carefully construct the full model. A comparative study of multivariate Granger networks \cite{Siggiridou_Comparison_2019} indicates that the Restricted conditional Granger causality index (RCGCI) from \cite{GrangerMultivariate_Siggiridou} is the most suitable Granger causality estimation scheme for the financial data analysed in this article. At the heart of RCGCI lies the construction of the full model by starting with an empty regression model and sequentially adding variables $X_{t-k\tau}^{(i)}$ with a lag of $k$ time units to it, if they reduce the \textit{BIC} of the regression \cite{Schwarz1978}. Hence, the resulting full model may not contain lagged representations of all possible explanatory variables but only some selected $\left( X^{(i_1)},\dots, X^{(i_I)}\right)$ and therefore guarantees sparsity to prevent overfitting in the estimation process. For those variables, \textit{CGC} can be computed by removing them from the full model and fitting the reduced model, whereas the remaining selected variables are conditioned on as the background information $Z$. For the other $X_j$ which have not been included in the full model, the \textit{CGC} is set to zero as no causal relationship had been estimated.

\subsubsection{Details on the Estimation}
Because the financial returns analysed in this article are known to have an almost nonexistent autocorrelation \cite{MantegnaStanleyBook} and to avoid overfitting, we restrict our models to maximum lags of one time step. As this represents a full day of trading activity, data with a lag of two time units (i.e. two days) in the fast-paced and constantly adapting environment of financial markets yields little additional contribution to the full Granger model as shown by an exploratory analysis. Previous studies have shown that principle component analysis (PCA) can be used to distinguish between noise and collective effects in financial time series \cite{RandomMatrixStanley,RandomMatrixLalouxPotters}. Hence, we perform PCA on the raw data, only keep the principle components with the largest eigenvalues so that their sum describes $90\%$ of the total variation in the data and discard the remaining principle components as noise before performing the inverse transformation back into the original feature space. Note that the sparsity of the RCGCI algorithm also limits the influence of noise on the results. Averaging over all sectors and all time periods under consideration, the typical ratio between the variance explained by the full regression model and the variance of the data is $\sigma_F^2 /\sigma_\text{Data}^2 \approx 96\%$ indicating a good fit of full the regression model and a high signal to noise ratio.

\subsection{Helmholtz-Hodge-Kodaira Decomposition}
The reconstructed network of the causality flux  between multivariate time series might not be ad hoc trivial to interpret. Circular causalities ($A$ causes $B$, $B$ causes $C$ and $C$ causes $A$) can be present and inspecting the network with the naked eye may not be sufficient to understand its structure. The Helmholtz-Hodge-Kodaira Decomposition (HHKD) is a tool to analyse the flux in networks and to disentangle the flow into upstream and downstream directions \cite{Johnson_HodgeTutorial2013,Strang_PhDThesis}.\\

\subsubsection{Mathematical Formulation of the Unidirectional HHKD}

The Helmholtz decomposition theorem states that any well-behaved vector field $\textbf{F(r)}\in\mathbb{R}^n$ can be decomposed into two components $\textbf{F(r)}=\textbf{G(r)}+\textbf{R(r)}$, a gradient field \textbf{G(r)} and a divergence-free field \textbf{R(r)}. The rotation-free field \textbf{G(r)} can be expressed as the gradient of a potential $\textbf{G(r)} = -\nabla_\textbf{r} \Phi\textbf{(r)}$ such that the potential determines the direction of a flux in the space of \textbf{r}. The divergence-free or solenoidal field $\textbf{R(r)}$ has the property that no point $r$ is a source or sink of the observed flux as $\forall \textbf{r}: \nabla_\textbf{r} \cdot \textbf{R(r)}=0$. Note that a third component can exist and represents a background flux into and out of the system, but is usually ignored as one assumes that the system of interest is sufficiently closed.\\

\begin{figure}
    \centering
\includegraphics[width=0.75\textwidth]{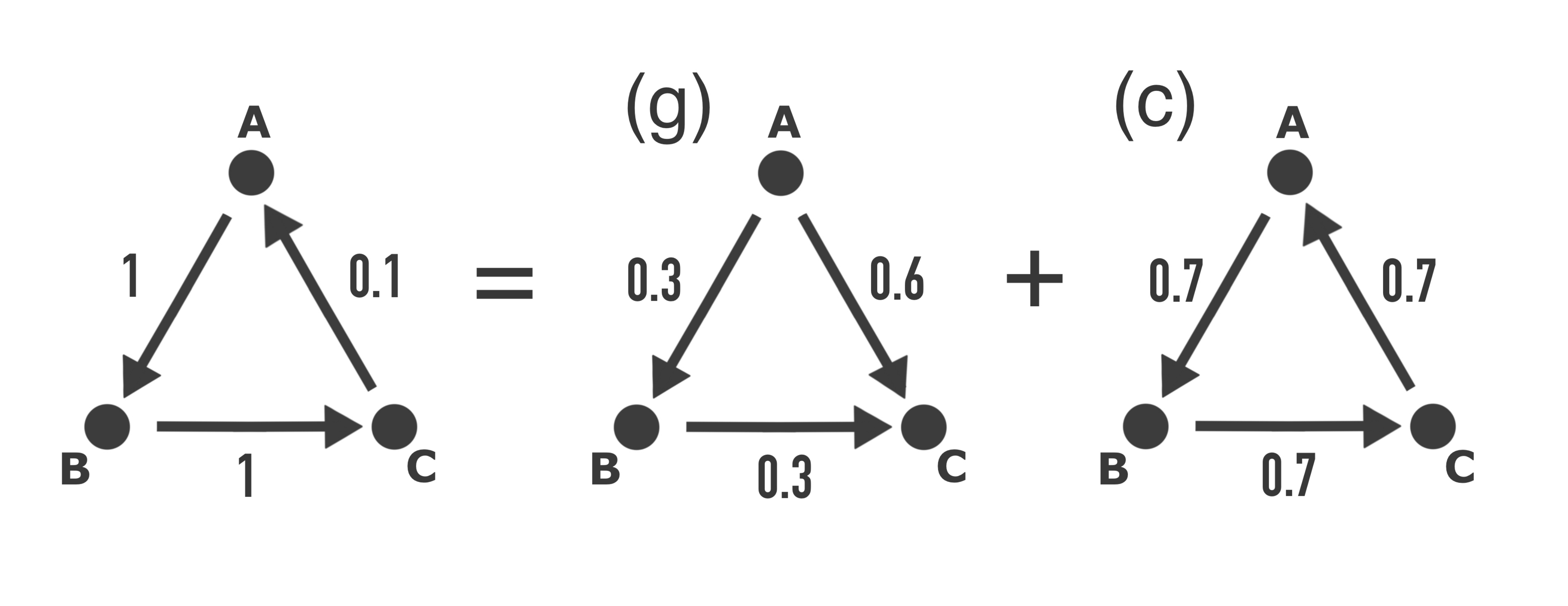}
\caption{Example of the Helmholtz-Hodge-Kodaira decomposition for a single graph into a gradient-based graph (g) and a circular graph (c). Note that direction of the flux between A and C is different in (g) and (c) which is the same as changing the sign $J^{(g)}_{AC} = - J^{(g)}_{CA}$ and hence their sum is given by $ J^{(g)}_{CA} + J^{(c)}_{CA} = -0.6+0.7 = 0.1$ and reconstructs the original flux $J_{CA}$. Also note that the total flux between two nodes is path independent for (g) as $J^{(g)}_{AC} = J^{(g)}_{AB}+J^{(g)}_{BC}$.}
\label{fig:HHKD_example}
\end{figure}

The same reasoning can be applied to a flow network on a discrete graph \cite{Johnson_HodgeTutorial2013,Strang_PhDThesis}. Let $J_{ij}$ be the observed flow from node $i$ to $j$ with the antisymmetric property $J_{ij} = -J_{ji}$. It can be shown that a unique decomposition $J_{ij} = J_{ij}^{(g)} + J_{ij}^{(c)}$ exists such that $J_{ij}^{(g)}$ is the gradient and $J_{ij}^{(c)}$ the circular flow from $i$ to $j$. In this decomposition, the gradient flux fulfils $J_{ij}^{(g)} = G_{ij}\left( \Phi_i - \Phi_j \right)$ for some background potential $\Phi$ assigned to each node and with the standard choice for weights between two nodes being $G_{ij}=1$. The circular flow fulfils $\forall i: \sum_j J_{ij}^{(c)} \overset{!}{=}0$, i.e. that for each node the total influx is equal to its total outflow. For a simple network with three nodes, this decomposition is illustrated in figure \ref{fig:HHKD_example}. The potential and its associated gradient flow can be obtained from the least square estimation

 \begin{equation}
 \label{eq:MinimisationHHKD}
        \underset{\textbf{J}^{(g)}}{\min} \left( I\right) \hspace{0.2cm} \textmd{ with } \hspace{0.2cm}  I= \frac{1}{2}\sum_{i<j} \frac{1}{G_{ij}} \left(J_{ij} - J_{ij}^{(g)} \right)^2  = \frac{1}{2}\sum_{i<j} \frac{1}{G_{ij}} \left(J_{ij} - G_{ij}\left( \Phi_i - \Phi_j \right) \right)^2
\end{equation}
and the circular flow is then simply the difference $J_{ij}^{(c)} =J_{ij}- J_{ij}^{(g)}$. For the standard choice $G_{ij}=1$, this formulation also has the useful property that the net gradient flux is the same along all paths between any two nodes. Because the gradient flow only depends on the potential difference $\Phi_i - \Phi_j$, the same gradient flow can also be obtained if the potentials have a constant offset $\Phi_i \rightarrow \Phi_i + \Phi_\mathcal{O}$. Hence, the minimisation of equation~\eqref{eq:MinimisationHHKD} needs an additional constraint to produce unique results for $\Phi$ such as $\Phi_n = 0$ or $\sum_i \Phi_i= 0$.

\unskip

\subsubsection{Bidirectional Flows}
Whether it be through noise or feedback loops, it is in general possible that the RCGCI algorithm estimates that $CGC_{X\rightarrow Y}>0$ and $CGC_{Y\rightarrow X}>0$, i.e. that two time series are estimated to Granger cause each other and that the flux cannot be defined as antisymmetric $J^{(b)}_{ij} \neq -J^{(b)}_{ji}$ where the superscript $b$ denotes the bidirecionality and $J^{(b)}_{ij}\geq 0$. Naively, computing the net flow $CGC_{X\rightarrow Y} - CGC_{Y\rightarrow X}$ seems like a reasonable choice, but this discards the information about the relative strength of the net flux. Consider a system in which $CGC_{A\rightarrow B}=0.6$, $CGC_{B\rightarrow A}=0.1$, $CGC_{C\rightarrow D}=5.5$, $CGC_{D\rightarrow C}=5$. The net flux $A\rightarrow B$ and $C\rightarrow D$ is $0.5$, but scaling this with the total flux between the node pairs shows that this difference is much less significant for the flux $C\rightarrow D$.\\

A bidirectional version of the HHKD that reflects these considerations is presented in \cite{Kichikawa2021}. The authors argue to interpret $G_{ij}$ as an analogy to the conductance in electrical circuits and that a high flux between both nodes should correspond to a high conductance whereas a low total flux indicates a high resistance. Hence, they propose to split the original bidirectional network into two graphs which are then used to perform the HHKD: The difference of the flux in both directions between two nodes in the bidirectional network $J^{(b)}$ is defined as the net flux $J_{ij}\coloneqq J^{(b)}_{ij}-J^{(b)}_{ji} $ and forms a unidirectional network with the antisymmetry $J_{ij} = -J_{ji}$ to which the HHKD can be applied. The sum of the absolute values of the flux in both directions is used as the conductivity $G_{ij}\coloneqq J^{(b)}_{ij}+J^{(b)}_{ji} $ which is the same in both directions $G_{ij} = G_{ji}$. The minimisation in equation \eqref{eq:MinimisationHHKD} can then be applied to the two networks $(J,G)$ to receive the HHKD ranking of the nodes.\\
This generalisation also gives rise to a helpful interpretation of the potential differences of the nodes: Consider only the flux between two nodes $i$ and $j$ isolated from the rest of the graph.  Let $J_{ij}$ be the net flow from $i$ to $j$ and $G_{ij}$ the total flow. The contribution of this connection to the functional $I$ which is minimised is then given by   
\begin{equation} 
\label{eq:HHKD_OnlyTwoNodes}
I_{ij} =\frac{1}{2} \frac{1}{G_{ij}} \Big(J_{ij}-J_{ij}^{(g)}\Big)^2 = \frac{1}{2} \frac{1}{G_{ij}} \Big(J_{ij}-G_{ij}\left(\Phi_i -\Phi_j \right)\Big)^2 = \frac{1}{2} \frac{1}{G_{ij}} \Big(J_{ij}-G_{ij}\Delta_{ij}\Big)^2
\end{equation}
where $\Delta_{ij}$ expresses the potential difference between the two nodes. Minimisation of $I_{ij}$ with respect to $\Delta_{ij}$ leads to
\begin{equation}
    \frac{\partial I_{ij}}{\partial\Delta_{ij}} = -J_{ij} + G_{ij}\Delta_{ij} \overset{!}{=}0 \Leftrightarrow \Delta_{ij} = \frac{J_{ij}}{G_{ij}} \equiv \frac{\textmd{Net Flow}}{\textmd{Total Flow}}.
\end{equation}
Hence, this rule-of-thumb approximation which disregards all other edges shows that the potential difference can be interpreted as the ratio between net and total flow between the two nodes. In particular, if the flux in one direction is much larger than in the other direction $J^{(b)}_{ij} \gg J^{(b)}_{ji}$, then net and total flow are almost identical $J_{ij} \approx G_{ij}$ so that $ \Delta_{ij} \approx 1$. Therefore, as described in \cite{Kichikawa2021}, one unit of potential difference can be interpreted as a separation of approximately one layer between the nodes $i$ and $j$.

\subsubsection{Circularity and Hierarchy}

Once the flow network is decomposed into gradient-based and circular flux, one can compare their respective contributions to the net flux~\cite{Kichikawa2021,HHKD_GammaLambda,HHKD_GammaLambda2}. It is possible to quantify the contribution of the gradient-based flux via the $L^2$ norm as
\begin{equation}
\label{eq:Gamma}
    \Gamma =  \frac{1}{2}\sum_i \Gamma_i =  \frac{1}{2}\sum_i \sum_j G_{ij}\left( J^{(g)}_{ij}\right)^2
\end{equation}
and that of the circular flux as 
\begin{equation}
\label{eq:Lambda}
    \Lambda =  \frac{1}{2}\sum_i \Lambda_i  = \frac{1}{2}\sum_i \sum_j G_{ij}\left( J^{(c)}_{ij}\right)^2
\end{equation}
where $\Gamma_i$ and $\Lambda_i$ denote the contribution of the respective $i^\textmd{th}$ node. Normalising them with the total flux 
\begin{equation}
\label{eq:totalflux_N}
    N = \frac{1}{2}\sum_i \sum_j G_{ij}\left( J_{ij}\right)^2
\end{equation}
leads to the definition of
\begin{equation}
\label{eq:gamma_lambda_gradient_circular}
    \gamma = \frac{\Gamma}{N} \hspace{0.2cm}\textmd{and}\hspace{0.2cm} \lambda = \frac{\Lambda}{N}
\end{equation}
which fulfil $\gamma + \lambda = 1$. A completely hierarchical network will have $\gamma = 1$, a completely circular network $\lambda = 1$. A high $\gamma \gg \lambda$ indicates that the underlying potential and its corresponding hierarchy have been cleansed from noise and insignificant loops and now accurately reflects the true structure of the underlying dynamics.

\subsection{Test on Synthetic Data}

\begin{figure}
\centering
\includegraphics[width=0.99\linewidth]{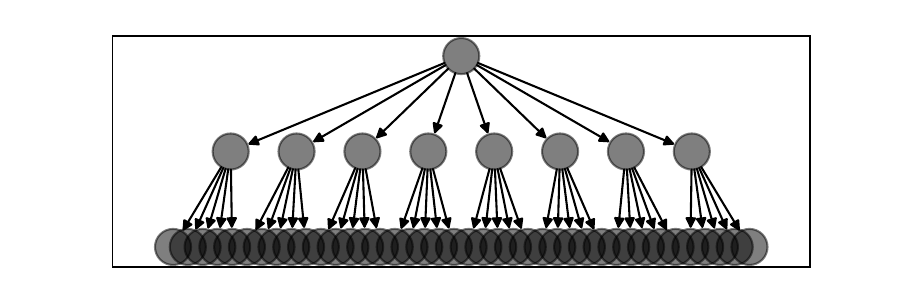}
\caption{The network structure used for the vector autoregression which generates synthetic time series. One node is at the top of the hierarchy without any causal parent whereas 8 nodes are in the second layer and 40 are in the final layer. Each node in the second layer is the parent node of 5 nodes in the final layer and have the node in the first layer as their parent node. Sketched via the software \cite{NetworkX}.}
\label{fig:SyntheticNetwork}
\end{figure}

To test the pipeline of RCGCI and HHKD, we simulate 50 realisation of a network of 49 time series with the network structure given in figure \ref{fig:SyntheticNetwork}. Because the RCGCI-HHKD pipeline is supposed to uncover the hierarchy of time series, the synthetic network in \ref{fig:SyntheticNetwork} was chosen to represent a hierarchical structure. Each node $X^{(i)}$ of the network is simulated for 250 time steps in a vector autoregression according to 
\begin{equation}
\label{eq:Synthetic_VAR}
    X^{(i)}_{t+1} = -0.5 X^{(i)}_t - 0.5X^{pa(i)}_t + \epsilon
\end{equation}
where $pa(i)$ denotes the parent node of $i$ (if it exists) and $\epsilon \overset{iid}{\sim}\mathcal{N}(0,\sigma)$ with $\sigma = 1$. Note that 49 time series for 250 time steps correspond to one year of trading day data in the database described in section \ref{sec:DataFrench} and that the standard deviation $\sigma = 1$ is roughly equal to the standard deviation of the data. Hence, these synthetic time series provide a realistic artificial version of the observed data. In the artificial network, the nine nodes with index $\mathcal{I}=(0,\dots,8)$ in the first and second layer have a heightened position in the causal hierarchy. The HHKD is used on the Granger causality network estimated by RCGCI and it is evaluated how many of the top nine nodes in the estimated hierarchy actually belong to the set $\mathcal{I}$. For the 50 realisations of the network, the average of the ensemble of detection rates is $94\%$ and the median is even $100\%$. Though not shown here, for synthetic cyclic networks, the RCGCI also successfully estimates the loop topology. Hence, the combination of RCGCI and HHKD accurately estimates the network structure. Even after adding observation noise $\mathcal{N}(0,\sigma_\text{obs})$ to the simulated data, the top nodes of the networks hierarchy are still estimated with a high accuracy beyond the random expectation for noise up to $\sigma_\text{obs}\approx1$.

\section{Results} 
\label{sec:hhkd_results}

\subsection{Year by Year}
\begin{figure}
    \centering
    \includegraphics[width=0.85\linewidth]{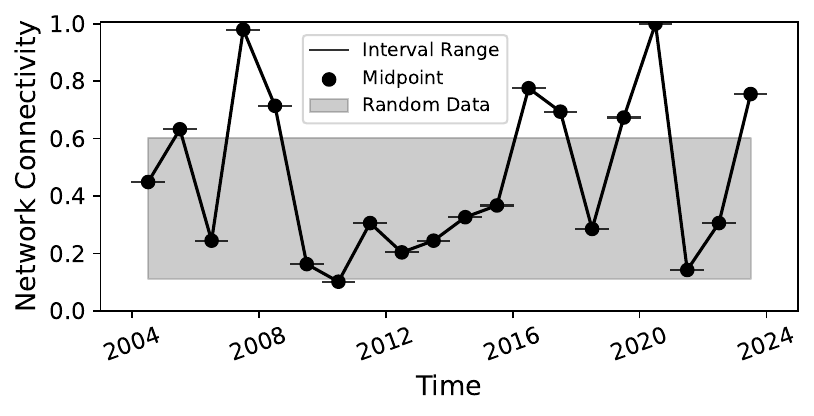}
    \caption{Results of the RCGCI-HHKD analysis for annual data from \cite{FrenchData}. The grey shaded area is the CI for the network connectivity of random data without any causal coupling.}
    \label{fig:AnnualDisjoint_HHKD}
\end{figure}

To gain the most insight from the HHKD, it is necessary to have a complete network in which all sectors are included. By defining the network connectivity as the percentage of nodes which are coupled to the network, the RCGCI-HHKD analysis is performed on the annual data during the last 20 years from 2004 to 2023 to identify periods with perfect connectivity. Additionally, by taking a random subset of trading days $t_{i_1},\dots,t_{i_{250}}$ without any chronological ordering, it is possible to estimate the connectivity which might be erroneously estimated for causally independent data. 50 random subsets are drawn from the observed data, their network connectivities are estimated with the RCGCI-HHKD pipeline and the respective confidence interval (CI) is calculated.\\
The results in figure~\ref{fig:AnnualDisjoint_HHKD} show that the market mostly remains within the range expected from random time series, but some periods exhibit a spike to a significantly high level of network connectivity. It only reaches a perfect connectivity (i.e. that all sectors are coupled to the network) during the year 2020 and reaches a connectivity of almost $100\%$ (only one sector is decoupled) in 2007. Hence, the 2020 period will be the focus of the latter half of this section.\\
\unskip

To gain more insights into general structures of the RCGCI networks across the years, the sum of Granger causality influx and outflux is recorded for each of the 49 sectors as well as during how many years they are connected to the network by influx or outflux links. After normalising all of those quantities to the same scale, kernel density estimation (KDE, \cite{Scott2015kde}) shows that for inwards and outwards directions, the total flux and the linkage rate are distributed similarly. The influx has a highly peaked distribution which roughly resembles a Gaussian bell curve whereas the outflux has a higher variance and, notably, a fat tail at high values. Hence, while most sectors have a similar influx of Granger causality, some sectors drive the other sectors with a much stronger outwards Granger causality than most others. It is therefore more interesting to focus on the sectors with particularly high or low outflux of Granger causality. The sectors Rubbr (Rubber and Plastic Products), BldMt (Construction Materials), Mach (Machinery), Trans (Transportation) and, perhaps surprisingly, Banks show no outflux of Granger causality during any of the periods. Gold (Precious Metals) and, on the second position, Cnstr (Construction) have a much higher sum of Granger causality outflux and a much higher rate of outwards linkage than all other sectors. 

\begin{figure}
    \centering
    \includegraphics[width=0.7\linewidth]{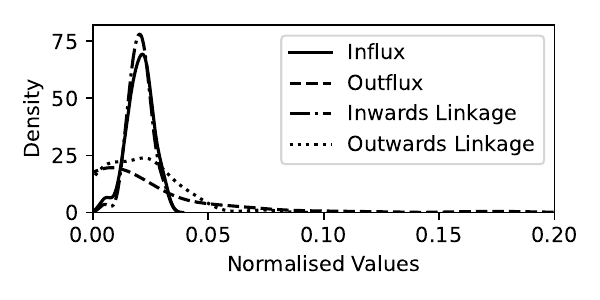}
    \caption{For the analysis of annual data from 2004 to 2023, KDE of the sum of all influx and outflux of Granger causality and of the total number of years with at least one inwards or outwards link in the RCGCI network. Values on the x-axis have been normalised to the same scale.}
    \label{fig:KDE_DisjointYears}
\end{figure}
\unskip

\subsection{Complete Graphs during the 2020 Covid Pandemic}
\label{sec:HHKD2020}

To further investigate the completely connected network for the year 2020, the RCGCI-HHKD pipeline is used to analyse time windows of 12 months which are shifted by one month and scan over the year 2020. This process starts with the time interval January 2019 to January 2020 and ends with the period from December 2020 to December 2021. Note that the figures displaying these results use the midpoint of each time period on the x-axis, e.g. July 2019 for the period from January 2019 up to January 2020. For each period with a complete network, the parameter $\gamma$ is calculated according to equation~\eqref{eq:gamma_lambda_gradient_circular} to quantify the contribution of the gradient flow to the observed flux. Because $\lambda =1-\gamma$, the calculation of $\lambda$ is omitted.\\

\begin{figure}
    \centering
    \includegraphics[width=0.9\linewidth]{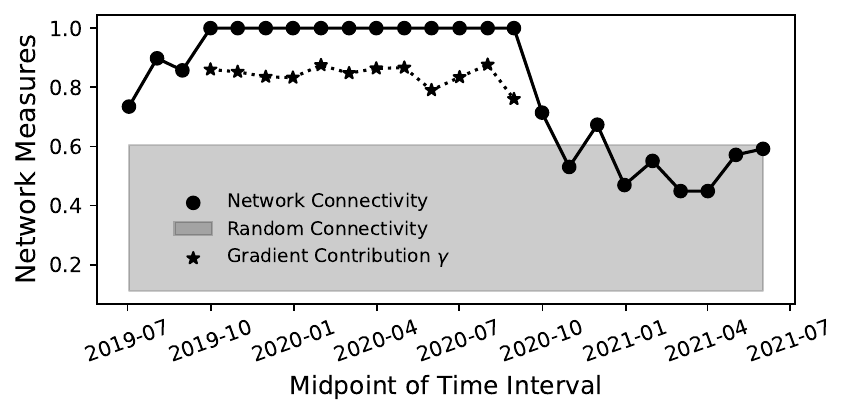}
    \caption{For time periods of 12 months, the network connectivity as the percentage of sectors connected to the network is displayed here against the random connectivity expected for independent time series. If the network is complete and has a connectivity of $1$, the gradient contribution $\gamma$ is also calculated according to equation~\eqref{eq:gamma_lambda_gradient_circular}. Note that the time on the x-axis is the midpoint of the 12 months intervals of data.}
    \label{fig:ConnectGammaLambda_HHKD_2020}
\end{figure}

Figure~\ref{fig:ConnectGammaLambda_HHKD_2020} shows the results for the connectivity and the gradient contribution $\gamma$. Whether the network is complete or not strongly depends on whether March 2020 is included in the data. During this period, the gradient contribution is typically around $\gamma \approx 0.8$ and therefore stronger than the rotational flow $\lambda$. Though due to the quadratic $L^2$ norm used to calculate \eqref{eq:gamma_lambda_gradient_circular}, a rotational component $\lambda \approx 0.2$ is nevertheless a non-negligible contribution to the total flow.
\unskip
\begin{figure}
    \centering
    \includegraphics[width=0.99\linewidth]{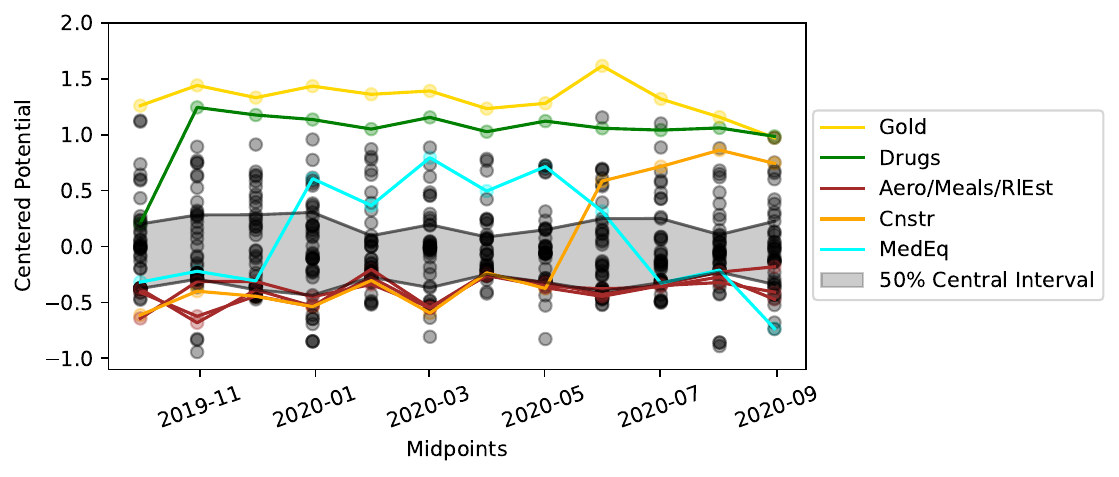}
    \caption{For the same time intervals as in figure~\ref{fig:ConnectGammaLambda_HHKD_2020},the potentials $\Phi_i$ of each sector are shown. Note that for each time interval, the potentials have been centered via $\sum_i \Phi_i= 0$. Some selected sectors are shown in colour and the grey area shows the spread between the $25\%$ and $75\%$ quantiles for each time period.}
    \label{fig:Covid2020Potentials}
\end{figure}
\unskip
Looking at the ensemble of Granger causality flux matrices for the 12 time windows with connectivity $1$ reveals that every single sector has always some causality influx (i.e. it is estimated to be Granger caused by another sector) for all 12 time windows with the exception of the sector Gold which only has an influx of Granger causality for 3 of the 12 periods. The sectors PerSv (Personal Services), Other, Aero (Aircraft) and Trans (Transportation) have no outflux of Granger causality during any of the 12 time windows and for the latter two sectors, this might reflect the travel restrictions imposed during this period. In contrast to the sum of Granger causality outflux for the disjoint long-term analysis in the previous section, Gold is no longer the sector with the highest total outflux ($\approx 5.00$) but is far overtaken by Drugs (Pharmaceutical Products; $\approx 35.8$), Hshld (Consumer Goods; $\approx 28.4$) and Cnstr ($\approx 10.6$) with some other sectors at a slightly higher level than Gold, too.\\

\begin{figure}
\centering
\includegraphics[width=1.0\linewidth]{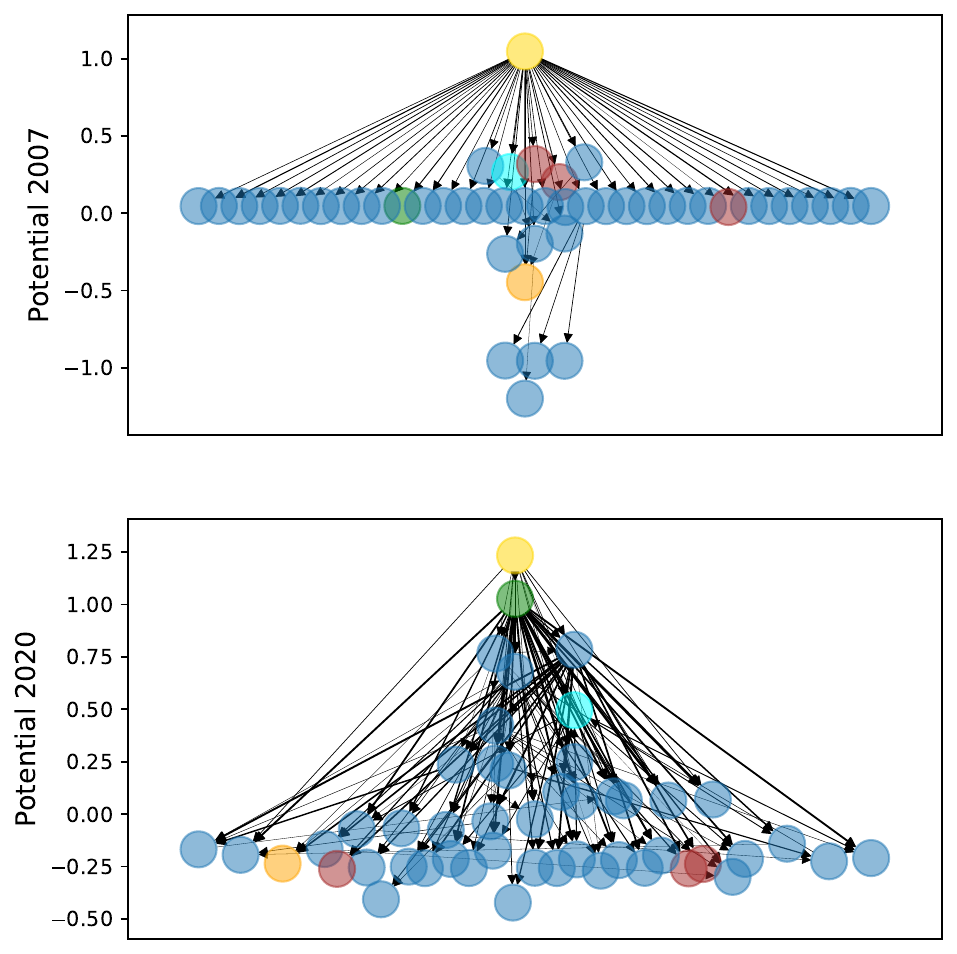}
\caption{The estimated Granger causality influence network ordered by the HHKD potentials for the periods from January 2007 to December 2007 (the sector FabPr is not shown because it has no link to any other sector) and from October 2019 to September 2020. The width of the arrows reflects strength of the Granger causality and selected sectors are highlighted with the same colour coding as in fig~\ref{fig:Covid2020Potentials}.}
\label{fig:NetworkVisualisations}
\end{figure}

For the periods with complete graphs, the potential $\Phi_i$ can be calculated for every single node. High $\Phi_i$ indicate a high position in the hierarchy of Granger causality and that the sector is a cause rather than an effect. Because of the large contribution $\gamma$ of the gradient flux shown in figure~\ref{fig:ConnectGammaLambda_HHKD_2020}, this hierarchy is not obstructed by strong circular fluxes in the system and indeed reflects the underlying dynamics. Figure~\ref{fig:Covid2020Potentials} shows the potentials for all 12 time windows with a complete graph. These potentials can be compared to each other because they have been normalised to fulfil $\sum_i \Phi_i = 0$ for each time period. The range between minimum and maximum potential values has a mean of $2.0$ with standard deviation $\pm0.2$ across these periods and therefore reflects a network of approximately 3 different levels with a fairly stable potential range. Additionally, for the period from October 2019 to September 2020, the full network is depicted in figure~\ref{fig:NetworkVisualisations} where the nodes' vertical positions reflect their potential values. Some selected sectors have been highlighted in these plots: The sectors Gold and, with the exception of the first interval, Drugs are consistently at the top of the hierarchy and their potentials have a low variance. Similarly, the potentials of the sectors Aero, Meals (Restaurants, Hotels, Motels) and RlEst (Real Estate) have some mean values and variances. Therefore, these sectors are consistently found at the bottom of the potential hierarchy.\\
Some other sectors have a high variance and change their position in the hierarchy rather drastically: The potential of Cnstr has the highest variance and this sector moves upwards in the hierarchy during the latter third of the periods. This might reflect the increase of construction material prices and their effect on the construction businesses and, as a cascading effect, on other business sectors during the beginning of 2021 \cite{web_ConstructionPriceIncrease}. The second highest variance is observed for the potential of MedEq (Medical Equipment). This sector rises in the hierarchy during the peak of Covid, which probably reflects the increasing demand for products such as face masks and testing equipment. The sudden decline of MedEq in the hierarchy starts in the time windows which already include the first weeks of widespread vaccinations in western countries which was interpreted as a sign of the end of the pandemic and hence of a lower importance of such equipment.\\
Because of the small but notable contribution of the rotational flux to the system during the Covid crisis, we also investigate which nodes have a strong rotational component $\Lambda_i$ as in equation~\eqref{eq:Lambda}. For each time period with a complete network graph, the values $\Gamma_i$ and $\Lambda_i$ are calculated according to equations~\eqref{eq:Gamma} and~\eqref{eq:Lambda} and the rotational component is normalised in two ways: $\Lambda_i^{(N)} = \Lambda_i / N$ denotes how much the rotational flux of node $i$ contributes to the total flux $N$ from equation~\eqref{eq:totalflux_N}. $\lambda_i = \frac{\Lambda_i}{\Lambda_i + \Gamma_i}$ denotes whether the flux of node $i$ is rather dominated by rotational flows or by the gradient flow. For each sector $i$, the mean values of $\Lambda_i^{(N)}$ and $\lambda_i$ are calculated over the time periods of consideration. While the sectors Drugs, Hshld and Cnstr have the highest, second highest and fourth highest mean value of $\Lambda_i^{(N)}$ and contribute much to the rotational flow, their own flow does not show a particularly high contribution $\lambda_i$ of the rotational component. Rather their $\Lambda_i^{(N)}$ is high because they in general have strong causality links to other sectors. The third and fifth highest values of $\Lambda_i^{(N)}$  are found for the sectors Toys (Recreation) and Softw (Computer Software) and they also have the second and third highest values of $\lambda_i$ with $\lambda_i\approx 0.47$ for both sectors. These sectors not only provide a strong rotational contribution to the total observed causality flow ($\Lambda_i^{(N)}$) but also experience an almost equally strong effect of the gradient and rotational flows ($\lambda_i$) on their own dynamics. This makes them interesting candidates for future research to better understand the circular dependencies in the financial network.
\unskip

\subsection{The 2007 Financial Crisis}
Because the Granger network for 2007 is not completely connected, it will not be analysed as deeply as the 2020 network in this manuscript. But since only a single sector (FabPr, Fabricated Products) is disconnected from the rest, it might be justified to briefly focus on the reduced network of the 48 connected sectors; not least because this period also coincides with the onset of the financial crisis in the late 2000s \cite{ChangePointsSP500} and serves as an interesting comparison to the Covid crisis. This reduced network is depicted in figure \ref{fig:NetworkVisualisations} and visual inspection shows a much more streamlined flow than for the network during 2020: Whereas the 2020 network is much more entangled, the 2007 network mostly consists of links from the sector Gold to other sectors with much fewer links between the other sectors, resulting in a shape reminiscent of the depictions of Aton in ancient Egyptian artworks. This is confirmed by the estimation of the gradient flow contribution for the reduced network which yields $\gamma^{(2007)} = 0.98$ and shows an even higher gradient contribution than for any of the 2020 networks. Although the 2007 Lehman crisis is generally known as the financial crisis, the financial sectors do not have a particularly important position in the hierarchy of Granger causality during this period and perhaps rather act as mediators of causality than as causal drivers. This surprising result might indicates that the causality analysis for this period does not fully represent the processes in the real economy but uncovers more subtle relationships between the time series.

\section{Discussion}
\label{sec:hhkd_discussion}
Analysis of the disjoint annual periods in figure~\ref{fig:AnnualDisjoint_HHKD} indicates that a highly connected Granger causality network coincides with market crises. The highest connectivity values close to $100\%$ were observed during the 2020 Covid crisis and the 2007 beginning of the financial crisis. Other spikes occur during the year 2023, possibly reflecting the collapse of several mid-size banks and the threat of a contagion in the US banking crisis \cite{Ozili2023}, and in 2016 following the market turmoil after the unexpected election of Donald Trump \cite{StockMarketReaction2016}. These results align with the causality estimation in \cite{GrangerCausality_FinancialNetworks} as well as with research on financial correlation matrices which also shows an increase in coupling between financial time series during times of crises \cite{Mnnix2012,Stepanov2015,Rinn2015DynamicsOQ}. The lower coupling during periods of a healthy market reflect that the time series are more independent and diversified which reduces the overall risk in the market.\\
The precious metal sector is usually at the top of the upstream in the Granger causality hierarchy but rather because of its own lack of causal drivers and not necessarily because of the influence it exerts on other sectors. Note that the returns of this sector are calculated based on the companies which trade precious metals and do not directly contain the prices of gold and other metals. Adding this might be an interesting endeavour for future work. During the Covid crisis, the high position of Drugs in the causality hierarchy as well as the rise of MedEq during the pandemic's peak reflect our intuition about the economy during the year 2020. Perhaps surprisingly, the financial sectors do not have a high position in the hierarchy of Granger causality and especially the banking sector is found rather far downstream. This might be interpreted as the financial companies being only a mediator of causal influence and providing the infrastructure to the flow of causality in the financial markets, but not actually driving the flow themselves. Our results therefore differ from the study in \cite{MacKay_network} where an analysis of the input-output network of business sectors shows that the energy and finance sectors have a high upstream position in the hierarchy. This is an important indicator that the financial market network analysed in our study does not simply resemble the real economy, but has its own dynamical behaviour. The high signal to noise ratio of the full regression models in the RCGCI algorithm and the strong contribution $\gamma\gg\lambda$ of the gradient flow indicates that the hierarchy estimated by the HHKD is reflective of the true underlying structure of the market dynamics. Interestingly, $\gamma$ was notably higher for the 2007 financial crisis than for the Covid crisis, possibly because the former was an endogenous crisis and the latter an external shock.\\
Numerous extensions can be made in future work to this project. Adding return time series of the precious metals' prices has already been suggested, but beyond this, macroeconomic variables like the inflation rate might be used as background variables $Z$ in equation~\eqref{eq:ConditionalGranger_FullModel}. Without attempting to make regression models to predict $Z$, these variables can still be used to calculate the Granger causality conditional on the macroeconomic information provided by them. This might be interpreted as the third translational component which is usually omitted in Helmholtz-Hodge considerations but represents an influx or outflux into the whole system of interest as discussed in~\cite{Strang_PhDThesis}. Moreover, the linear regression can be extended with interaction terms between two variables $X^{(i)}_{t-\tau} \cdot X^{(j)}_{t-\tau}$ or nonlinear functions \cite{ma2023linearnonlinearcausalityfinancial} to alleviate the shortcomings of Granger causality methods~\cite{Maziarz2015,Stokes}, but this might require larger amounts of data for reliable estimation and thus higher frequency data than the one available in~\cite{FrenchData}. Other methods from causal inference, such as the lead-lag relationship of complex Hilbert PCA \cite{Iyetomi2020,SoumaArticle} or transfer entropy \cite{CausalEntropyFinancialNetwork,Aste_DiMatteo_2017} can capture nonlinear nonlinear effects, but might require more data, too. Also, one could extend the RCGCI algorithm to include a bootstrapping procedure in the estimation of \eqref{eq:CGC} to get an uncertainty estimation of the Granger causality $CGC_{X\rightarrow Y}$. While the RCGCI and the standard formulation of Granger causality does not distinguish between positive and negative influences between variables similar to~\cite{StanleyCausality}, a multi-layer network approach could be used to separate the causal couplings based on their sign. However, extending the HHKD to multi-layer networks is required to evaluate this, perhaps based on the approach in~\cite{Helmholtz_multilayer}.\\
Because of the high network connectivity during crises, the RCGCI-HHKD pipeline is especially useful to describe the system dynamics during such periods. In particular, understanding the flow of causality and identifying the causal drivers during a crisis might allow policymakers to more effectively intervene to stop the crisis by focusing on the sectors which are upstream in the causality hierarchy. This could open up the possibility to stabilise the market with a minimally invasive intervention.\\
Finally, though beyond the scope of this work, we believe that the HHKD can help to overcome the limitations of the causality framework described by Judea Pearl~\cite{CausalPrimer}. Pearls approach to causality relies on directed acyclic graphs (DAG) between the variables and therefore requires an interaction network without any closed loops. While this is not always present in real systems, an adaptation of the HHKD might provide a suitable tool to extract such DAGs from real-world systems as the gradient component of the original graph.

\section*{Data and Code Availability}
The data used for this research project as well as the main code for the RCGCI-HHKD pipeline is published at~\cite{zenodo_hhkd} on zenodo.

\section*{Acknowledgments}
T.W. and H.I. would like to thank Wataru Souma (Rissho University) for valuable discussions. This research was funded by the Japan Society for the Promotion of Science with the Summer Program grant SP24309. T.W. is additionally supported by the German Academic Scholarship Foundation.

\section*{Abbreviations}
The following abbreviations are used in this manuscript:\\

\noindent 
\begin{tabular}{@{}ll}
BIC & Bayesian Information Criterion\\
CI & Confidence Interval\\
DAG & Directed Acyclic Graphs\\
HHKD & Helmholtz-Hodge-Kodaira Decomposition \\
KDE & Kernel Density Estimation\\
PCA & Principle Component Analysis\\
RCGCI & Restricted Conditional Granger Causality Index\\
\end{tabular}\\

\noindent 
Additionally, the following sector abbreviations were introduced by Ken French and are used in this article:\\
\noindent 
\begin{tabular}{@{}ll}
Aero & Aircraft\\
BldMt & Construction Materials\\
Cnstr & Construction\\
Drugs & Pharmaceutical Products\\
FabPr & Fabricated Products\\
Gold & Precious Metals\\
Hshld & Consumer Goods\\
Mach & Machinery\\
Meals & Restaurants, Hotels, Motels\\
MedEq & Medical Equipment \\
PerSv & Personal Services\\
RlEst & Real Estate\\ 
Rubbr & Rubber and Plastic Products\\
Softw & Computer Software\\
Trans & Transportation\\
Toys & Recreation
\end{tabular}

\bibliographystyle{unsrt}  
\bibliography{references}

\begin{thebibliography}{10}

\bibitem{Aldrich1995}
John Aldrich.
\newblock {Correlations Genuine and Spurious in Pearson and Yule}.
\newblock {\em Statistical Science}, 10(4), 1995.

\bibitem{GrangerCausality}
C.~W.~J. Granger.
\newblock Investigating causal relations by econometric models and cross-spectral methods.
\newblock {\em Econometrica}, 37(3):424--438, 1969.

\bibitem{CausalPrimer}
Judea Pearl, Madelyn Glymour, and Nicholas~P. Jewell.
\newblock {\em Causal Inference in Statistics: A Primer}.
\newblock Wiley, 2016.

\bibitem{Causality_Attractor_Lorenz}
R.~Quian Quiroga, J.~Arnhold, and P.~Grassberger.
\newblock Learning driver-response relationships from synchronization patterns.
\newblock {\em Phys. Rev. E}, 61:5142--5148, May 2000.

\bibitem{Causality_Ecosystems_Attractor}
George Sugihara, Robert May, Hao Ye, Chih-hao Hsieh, Ethan Deyle, Michael Fogarty, and Stephan Munch.
\newblock Detecting causality in complex ecosystems.
\newblock {\em Science}, 338(6106):496--500, 2012.

\bibitem{Causal_Inference_Optimal_Entropy}
Jie Sun, Dane Taylor, and Erik~M. Bollt.
\newblock Causal network inference by optimal causation entropy.
\newblock {\em SIAM Journal on Applied Dynamical Systems}, 14(1):73--106, 2015.

\bibitem{Aste_DiMatteo_2017}
Tomaso Aste and T.~Di~Matteo.
\newblock Sparse causality network retrieval from short time series.
\newblock {\em Complexity}, 2017:1–13, 2017.

\bibitem{GrangerMultivariate_Siggiridou}
Elsa Siggiridou and Dimitris Kugiumtzis.
\newblock Granger causality in multivariate time series using a time-ordered restricted vector autoregressive model.
\newblock {\em IEEE Transactions on Signal Processing}, 64(7):1759--1773, 2016.

\bibitem{Siggiridou_Comparison_2019}
Elsa Siggiridou, Christos Koutlis, Alkiviadis Tsimpiris, and Dimitris Kugiumtzis.
\newblock {Evaluation of Granger Causality Measures for Constructing Networks from Multivariate Time Series}.
\newblock {\em Entropy}, 21(11):1080, November 2019.

\bibitem{Iskrzyski2021}
Mateusz Iskrzyński, Freek Janssen, Francesco Picciolo, Brian Fath, and Franco Ruzzenenti.
\newblock Cycling and reciprocity in weighted food webs and economic networks.
\newblock {\em Journal of Industrial Ecology}, 26(3):838–849, December 2021.

\bibitem{Johnson_HodgeTutorial2013}
James~L. Johnson and Tom Goldring.
\newblock {Discrete Hodge Theory on Graphs: A Tutorial}.
\newblock {\em Computing in Science \& Engineering}, 15(5):42–55, September 2013.

\bibitem{Strang_PhDThesis}
Alexander Strang.
\newblock {\em Applications of the Helmholtz-Hodge Decomposition to Networks and Random Processes}.
\newblock Phd thesis, Case Western Reserve University, Cleveland, Ohio, August 2020.
\newblock Available at \url{https://case.edu/math/thomas/Strang-Alexander-2020-PhD-thesis-final.pdf}.

\bibitem{MantegnaStanleyBook}
Rosario Mantegna and H.~Stanley.
\newblock {\em An Introduction to Econophysics}.
\newblock Cambridge University Press, Cambridge, 2000.

\bibitem{Fujiwara2021}
Yoshi Fujiwara and Rubaiyat Islam.
\newblock Bitcoin’s crypto flow network.
\newblock In {\em Proceedings of Blockchain in Kyoto 2021 (BCK21)}. Journal of the Physical Society of Japan, November 2021.

\bibitem{Kichikawa2021}
Yuichi Kichikawa, Hiroshi Iyetomi, and Yuichi Ikeda.
\newblock {\em Who Possesses Whom in Terms of the Global Ownership Network}, pages 143--190.
\newblock Springer Singapore, Singapore, 2021.
\newblock In \textit{Big Data Analysis on Global Community Formation and Isolation: Sustainability and Flow of Commodities, Money, and Humans}, edited by Yuichi Ikeda, Hiroshi Iyetomi and Takayuki Mizuno.

\bibitem{WorldTradeWeb_PhysRevE.68.015101}
Ma~\'Angeles Serrano and Mari\'an Bogu\~n\'a.
\newblock Topology of the world trade web.
\newblock {\em Phys. Rev. E}, 68:015101, Jul 2003.

\bibitem{HandbookGraphs}
S.~Bornholdt and H.~G. Schuster, editors.
\newblock {\em Handbook of Graphs and Networks}.
\newblock Wiley-VCH, Weinheim, 2005.

\bibitem{LargeScaleStructure}
G.~Caldarelli and A.~Vespignani, editors.
\newblock {\em Large Scale Structure and Dynamics of Complex Networks}.
\newblock World Scientific Publishing, Singapore, 2007.

\bibitem{CausalEntropyFinancialNetwork}
Leonidas Sandoval.
\newblock Structure of a global network of financial companies based on transfer entropy.
\newblock {\em Entropy}, 16(8):4443--4482, 2014.

\bibitem{GrangerCausality_FinancialNetworks}
Angeliki Papana, Catherine Kyrtsou, Dimitris Kugiumtzis, and Cees Diks.
\newblock {Financial networks based on Granger causality: A case study}.
\newblock {\em Physica A: Statistical Mechanics and its Applications}, 482:65--73, 2017.

\bibitem{StanleyCausality}
Stavros~K. Stavroglou, Athanasios~A. Pantelous, H.~Eugene Stanley, and Konstantin~M. Zuev.
\newblock Hidden interactions in financial markets.
\newblock {\em Proceedings of the National Academy of Sciences}, 116(22):10646--10651, 2019.

\bibitem{FrenchData}
{Ken French}.
\newblock Us research returns data. 49 industry portfolios [daily].
\newblock \url{https://mba.tuck.dartmouth.edu/pages/faculty/ken.french/data_library.html}, 2024.
\newblock Continuously updated. Accessed: July 2024.

\bibitem{Fama2023}
Eugene~F. Fama and Kenneth~R. French.
\newblock {Production of U.S. SMB and HML in the Fama-French Data Library}.
\newblock {\em SSRN Electronic Journal}, 2023.

\bibitem{Chan2007}
Louis~K.C. Chan, Josef Lakonishok, and Bhaskaran Swaminathan.
\newblock Industry classifications and return comovement.
\newblock {\em Financial Analysts Journal}, 63(6):56–70, November 2007.

\bibitem{FAMA1997153}
Eugene~F. Fama and Kenneth~R. French.
\newblock Industry costs of equity.
\newblock {\em Journal of Financial Economics}, 43(2):153--193, 1997.

\bibitem{OverfittingBabyak}
Michael Babyak.
\newblock What you see may not be what you get: A brief, nontechnical introduction to overfitting in regression-type models.
\newblock {\em Psychosomatic medicine}, 66 (3):411--21, 05 2004.

\bibitem{Schwarz1978}
Gideon Schwarz.
\newblock Estimating the dimension of a model.
\newblock {\em The Annals of Statistics}, 6(2), March 1978.

\bibitem{RandomMatrixStanley}
Vasiliki Plerou, Parameswaran Gopikrishnan, Bernd Rosenow, Lu\'{\i}s~A. Nunes~Amaral, and H.~Eugene Stanley.
\newblock Universal and nonuniversal properties of cross correlations in financial time series.
\newblock {\em Phys. Rev. Lett.}, 83:1471--1474, 8 1999.

\bibitem{RandomMatrixLalouxPotters}
Laurent Laloux, Pierre Cizeau, Jean-Philippe Bouchaud, and Marc Potters.
\newblock Noise dressing of financial correlation matrices.
\newblock {\em Phys. Rev. Lett.}, 83:1467--1470, 8 1999.

\bibitem{HHKD_GammaLambda}
Taichi Haruna and Yuuya Fujiki.
\newblock Hodge decomposition of information flow on small-world networks.
\newblock {\em Frontiers in Neural Circuits}, 10, 2016.

\bibitem{HHKD_GammaLambda2}
Yuuya Fujiki and Taichi Haruna.
\newblock Hodge decomposition of information flow on complex networks.
\newblock In {\em Proceedings of the 8th International Conference on Bio-inspired Information and Communications Technologies (formerly BIONETICS)}, 01 2015.

\bibitem{NetworkX}
Aric Hagberg, Pieter Swart, and Daniel S~Chult.
\newblock Exploring network structure, dynamics, and function using {NetworkX}.
\newblock Technical report, Los Alamos National Lab.(LANL), Los Alamos, NM (United States), 2008.

\bibitem{Scott2015kde}
David~W. Scott.
\newblock {\em Multivariate Density Estimation - Theory, Practice, and Visualization}.
\newblock John Wiley \& Sons, New York, 2015.

\bibitem{web_ConstructionPriceIncrease}
Greg Zimmerman.
\newblock Construction materials prices increase more than 20 percent, 2022.
\newblock Available on \url{https://www.facilitiesnet.com/designconstruction/tip/Construction-Materials-Prices-Increase-More-Than-20-Percent--49437} and accessed on 24th July 2024.

\bibitem{ChangePointsSP500}
Martin Heßler, Tobias Wand, and Oliver Kamps.
\newblock {Efficient Multi-Change Point Analysis to decode Economic Crisis Information from the S\&P500 Mean Market Correlation}, 2023.

\bibitem{Ozili2023}
Peterson~K Ozili.
\newblock Causes and consequences of the 2023 banking crisis.
\newblock {\em SSRN Electronic Journal}, 2023.

\bibitem{StockMarketReaction2016}
Delia Diaconaşu, Seyed Mehdian, and Ovidiu Stoica.
\newblock {The Global Stock Market Reactions to the 2016 U.S. Presidential Election}.
\newblock {\em Sage Open}, 13(2):21582440231181352, 2023.

\bibitem{Mnnix2012}
Michael~C. M\"{u}nnix, Takashi Shimada, Rudi Sch\"{a}fer, Francois Leyvraz, Thomas~H. Seligman, Thomas Guhr, and H.~Eugene Stanley.
\newblock Identifying states of a financial market.
\newblock {\em Scientific Reports}, 2(1), September 2012.

\bibitem{Stepanov2015}
Yuriy Stepanov, Philip Rinn, Thomas Guhr, Joachim Peinke, and Rudi Schäfer.
\newblock Stability and hierarchy of quasi-stationary states: financial markets as an example.
\newblock {\em Journal of Statistical Mechanics: Theory and Experiment}, 2015(8):P08011, 8 2015.

\bibitem{Rinn2015DynamicsOQ}
Philip Rinn, Yuriy Stepanov, Joachim Peinke, Thomas Guhr, and Rudi Schäfer.
\newblock Dynamics of quasi-stationary systems: Finance as an example.
\newblock {\em {EPL} (Europhysics Letters)}, 110(6):68003, 6 2015.

\bibitem{MacKay_network}
R.~S. MacKay, S.~Johnson, and B.~Sansom.
\newblock How directed is a directed network?
\newblock {\em Royal Society Open Science}, 7(9):201138, 2020.

\bibitem{ma2023linearnonlinearcausalityfinancial}
Haochun Ma, Davide Prosperino, Alexander Haluszczynski, and Christoph Räth.
\newblock Linear and nonlinear causality in financial markets, 2023.

\bibitem{Maziarz2015}
Mariusz Maziarz.
\newblock {A review of the Granger-causality fallacy}.
\newblock {\em Journal of Philosophical Economics}, Volume VIII Issue 2(Articles), May 2015.

\bibitem{Stokes}
Patrick~A. Stokes and Patrick~L. Purdon.
\newblock {A study of problems encountered in Granger causality analysis from a neuroscience perspective}.
\newblock {\em Proceedings of the National Academy of Sciences}, 114(34):E7063--E7072, 2017.

\bibitem{Iyetomi2020}
Hiroshi Iyetomi.
\newblock {\em Collective Phenomena in Economic Systems}, page 177–201.
\newblock Springer Singapore, 2020.

\bibitem{SoumaArticle}
Wataru Souma.
\newblock Characteristics of principal components in stock price correlation.
\newblock {\em Frontiers in Physics}, 9:602944, 04 2021.

\bibitem{Helmholtz_multilayer}
Alp Kustepeli.
\newblock {On the Helmholtz Theorem and Its Generalization for Multi-Layers}.
\newblock {\em Electromagnetics}, 36(3):135--148, 2016.

\bibitem{zenodo_hhkd}
Tobias Wand.
\newblock {Helmholtz-Hodge-Kodaira Decomposition on Financial Data by Ken French}, 2024.
\newblock Available on Zenodo at \url{https://zenodo.org/doi/10.5281/zenodo.13340981}.

\end{thebibliography}

\end{document}